\documentclass[journal]{IEEEtran}
		
\newcommand{\E}{\mathrm{E}}

\usepackage{epsf,psfrag,amssymb,amsfonts,color,cite,fancybox}
\usepackage[mathscr]{eucal}
\usepackage[dvips]{graphicx}
\usepackage{caption}
\usepackage{subcaption}
\usepackage{ifpdf}
\usepackage{longtable}	
\usepackage{multicol}
\usepackage{float}
\usepackage{epstopdf}
\usepackage{bm}
\usepackage{multirow}
\usepackage[scientific-notation=true]{siunitx}

\ifCLASSINFOpdf

\else

\fi

\usepackage[cmex10]{amsmath}
\begin{document}

\title{PMU-Based Estimation of Dynamic State Jacobian Matrix and Dynamic System State Matrix in Ambient Conditions}

\author{Xiaozhe Wang,~\IEEEmembership{Member,~IEEE,}
        Janusz Bialek,~\IEEEmembership{Fellow,~IEEE,}
        Konstantin Turitsyn,~\IEEEmembership{Member,~IEEE.}

\thanks{Xiaozhe Wang is with the Department of Electrical and Computer Engineering, McGill University, Montreal, QC  H3A 0G4, Canada. email: xiaozhe.wang2@mcgill.ca}
\thanks{Janusz Bialek is with Skolkovo Institute of Science and Technology (Skoltech), Moscow, Russia, email: J.Bialek@skoltech.ru.}
\thanks{Konstantin Turitsyn is with the Department of Mechanical Engineering, MIT, Cambridge, MA 02139, USA. email: turitsyn@mit.edu.}
\thanks{Work of XW was partially supported by McGill University Star-up fund. Work of KT was supported by Skoltech Initiative and the Ministry of Education and Science of Russian Federation (grant No. 14.615.21.0001, grant code: RFMEFI61514X0001). The authors thank ten anonymous referees for their feedback and Deepjyoti Deka for his invaluable comments.
}
}

\maketitle

\begin{abstract}
In this paper, a hybrid measurement- and model-based method is proposed which can estimate the dynamic state Jacobian matrix and the dynamic system state matrix in near real-time utilizing statistical properties extracted from PMU measurements. The proposed method can be used to detect and identify network topology changes that have not been reflected in an assumed network model.
Additionally, an application of the estimated system state matrix in online dynamic stability monitoring is presented.
\end{abstract}


\IEEEpeerreviewmaketitle

\section{Introduction}

Power system security analysis heavily relies on the knowledge of the power flow Jacobian matrix, dynamic state Jacobian matrix and the dynamic state matrix $A$. The relationship between the first two matrices has been discussed in \cite{Pai:1990}. Those matrices can be easily calculated assuming that the power system model is known and state estimation provides a coherent set of measurements. However, information about the network model may be corrupted while state estimation results may be strongly affected by bad data resulting in erroneous calculation of the Jacobian matrices and the state matrix. { Wide adoption of phasor measurement units (PMUs) in recent years has made it possible to validate the assumed system model and estimate the values of parameters. One approach is to develop a dynamic equivalent of the dynamical system from PMU measurements \cite{Bialek:2007}-\cite{Zali:2013}. Another approach is to use PMU measurements to estimate some of the parameters of the assumed system model. Specifically, the authors of \cite{Bialek:2014} proposed a method to estimate the values of generator damping and/or inertia constants using estimates of system modes calculated from PMU measurements. The authors of \cite{Chakrabortty:2016} utilized the modes extracted from PMUs to estimate system parameters including inter-area transmission line impedances, intra-area Thevenin reactances, inertia and damping of the aggregated synchronous generators. The authors of \cite{Huang:2009} applied an extended Kalman filter to estimate the generator parameters using PMU data. The authors of \cite{Zhou:2011} discussed both model reduction and calibration approach (i.e., parameter estimation) for power system dynamic model using PMU data.}
The authors of \cite{Demarco:2016} proposed a PMU-based method to estimate the largest singular value of inverse Jacobian matrix that serves as a voltage stability indicator.

{This paper follows a slightly different approach in that it uses statistical properties extracted from the time-series of PMU measurements of voltages and angles to estimate the dynamic state
Jacobian matrix and the whole dynamic state matrix $A$  directly, i.e., without knowing all the parameters of the assumed power system model necessary to calculate those matrices.} The most relevant work is \cite{Chen:2016} in which PMU measurements of slight fluctuations in nodal power injections, voltage magnitudes and phase angles were used to estimate the power flow Jacobian matrix using linear total least-squares estimation. However, they did not consider estimation of the whole dynamic state matrix $A$. In this paper, we assume that we have available PMU measurements at all the generator buses from which
we can calculate rotor angles and rotor speeds \cite{Zhou:2011}-\cite{Liu:2011} and that power system dynamics are excited by stochastic load variations \cite{Wangxz:2015}\cite{Crow:2013}-\cite{Hines:2015}. Then we develop a method that ingeniously utilizes the Lyapunov equation \cite{Hines:2015}\cite{Gardiner:2009} to calculate the dynamic state Jacobian matrix from the covariance matrix of measurements assuming that we know generators’ moment of inertias $M$. The method is not purely measurement-based as we assume the knowledge of inertias $M$ (which are usually well known) but, most importantly, we do not need any information about the network model (topology and parameters). Estimation of the whole system state matrix $A$ requires additionally that generator damping coefficients $D$ are known.

We also develop a methodology for estimation of damping $D$ if it is unknown or uncertain. However that requires knowledge of the network model, generator {electromotive force} (emf) and variances of load variations, all of which were not required for estimation of the Jacobian matrix.

The proposed methodology combines nicely the statistical properties extracted from the PMU measurements and the inherent generator physics, working as a grey box bridging the measurement and the model. Importantly, the estimation does not require exhaustive computational effort and can be done in near real-time.
The proposed methodology of estimating the dynamic state Jacobian matrix may have a number of possible applications.
For example, we can use it for model validation purposes. By comparing the estimated dynamic state Jacobian matrix calculated using the proposed methodology with the one obtained using an assumed network model, any undetected network model changes can be identified. Similarly, the methodology of estimating damping $D$ can be used to validate the assumed values of damping $D$.

Regarding usefulness of the estimated system state matrix, there are various applications well explored in the literature \cite{Chiang:book}-\cite{Bialek:book}. In this paper, we demonstrate how online estimation of the system state matrix allows one to estimate the critical eigenvalue (i.e., the {rightmost eigenvalue}) which can be directly used as a good measure of proximity to instability \cite{Kundur:1992}. The left and right eigenvectors of the critical eigenvalue may further be estimated to predict the response of the system and design emergency control measures \cite{Cutsem:book}\cite{Hill:1993}. {The estimated matrices may also facilitate online oscillation analysis and control, generation re-dispatch, congestion relief and other emergency control design.}

The rest of the paper is organized as follows. Section \ref{sectionmodel} briefly introduces the power system dynamic model and then elaborates the proposed hybrid measurement- and model-based method. Section \ref{casestudy} conducts case studies in three test systems to illustrate the feasibility and accuracy of the proposed method, and investigates the impacts of measurement noise and window length. Section \ref{sectionapplications} presents an application of the proposed method in online stability monitoring.

\section{the system model and the proposed methodology}\label{sectionmodel}
We consider the general power system dynamic model:
\begin{eqnarray}
\dot{\bm{x}}&=&\bm{f}({\bm{x},\bm{y}})\label{fast ode}\\
\bm{0}&=&\bm{g}({\bm{x},\bm{y}})\label{algebraic eqn}
\end{eqnarray}
Equation (\ref{fast ode}) describes dynamics of generators,
and (\ref{algebraic eqn}) describes the electrical transmission system and the internal static behaviors of passive devices. $f$ and $g$ are continuous functions, vectors $\bm{x}\in\mathbb{R}^{n_{\bm{x}}}$ and $\bm{y}\in\mathbb{R}^{n_{\bm{y}}}$ are the corresponding state variables (generator rotor angles, rotor speeds) and algebraic variables (bus voltages, bus angles) \cite{Wangxz:CAS}\cite{Wangxz:ISCAS2017}.

{ We consider small signal stability in this paper. As opposed to injection data, ambient data is obtained without introducing any extra disturbance when power systems operate around the steady state. It can be used for continuous monitoring unlike ring-down response that requires waiting for infrequent large disturbances such as short-circuits or generator trips.} In this paper, we focus on ambient oscillations around the steady state, dominated mainly by the dynamics of generator angles. Hence, we demonstrate the proposed technique using the classical generator model, which can be regarded as an equivalent generator model of aggregated generators \cite{Chow:2014}. Therefore (\ref{fast ode})-(\ref{algebraic eqn}) becomes \cite{Chiang:book}\cite{Bialek:book}\cite{Pai:2012}:

\begin{eqnarray}
\dot{\bm{\delta}}&=&\bm{\omega}\label{swing-1}\\
M\dot{\bm{\omega}}&=&\bm{P_m}-\bm{P_e}-{D}\bm{\omega}\label{swing-2}
\end{eqnarray}
with
\begin{equation}
P_{e_i}=\sum_{j=1}^{n}E_iE_j(G_{ij}\cos({\delta}_i-{\delta}_j)+B_{ij}\sin(\delta_i-{\delta}_j))\label{swing-3}
\end{equation}
Particularly,  $\bm{\delta}=[\delta_1,...,\delta_n]^T$ is a vector of generator rotor angles, $\bm{\omega}=[\omega_1,...,\omega_n]^T$ is a vector of generator rotor speeds {w.r.t. the synchronous speed}, $M=\mbox{diag}(M_1,...M_n)$ whose diagonal entries are the moment inertia constants, $D=\mbox{diag}(D_1,...D_n)$ whose diagonal entries are damping factors, $\bm{P_m}=[P_{m_1},...,P_{m_n}]^T$ is a vector of generators' mechanical power input, $\bm{P_e}=[P_{e_1},...,P_{e_n}]^T$ is a vector of generators' electrical power output, $E_i$ is the emf magnitude behind the transient reactance; $G_{ij}+jB_{ij}=Y_{ij} \angle{\phi_{ij}}$ is the $(i,j)$th entry of the reduced admittance matrix that includes generators’ impedances. Rotor saliency is neglected.

In this paper, we assume that system loads are experiencing Gaussian variation around base case loading, which is the most common assumption to model load variation\cite{Pal:2010}. In the power system model (\ref{swing-1})-(\ref{swing-3}), the system loads are modeled as constant impedances. {Note that more realistic load models, e.g. ZIP loads, may be incorporated, since the effect of the nonlinear loads can be reflected at the internal nodes of the generators as injected currents, and further reflected by adding additional terms in (\ref{swing-3}) \cite{Elahi:1983}}. Load variation reveals itself in the diagonal elements of the reduced admittance matrix as \cite{Crow:2013}\cite{Nwankpa:1992}:
\begin{equation}\label{Ybus}
  Y(i,i)=Y_{ii}(1+\sigma_i dW_i)\angle{\phi_{ii}}
\end{equation}
where $W$ is a standard Wiener process, and $\sigma_i^2$ denotes the variance of load variation. It should be noted that similar to the approach in \cite{Crow:2013}, we assume that both active and reactive powers at a bus vary with the same level of randomness; the power factor ({and subsequently $\phi_{ii}$}) remains unchanged. Therefore, the power system model incorporating random load variation can be represented as:
\begin{eqnarray}
\dot{\bm{\delta}}&=&\bm{\omega}\label{swingrandom-1}\\
M\dot{\bm{\omega}}&=&\bm{P_m}-\bm{P_e}-{D}\bm{\omega}-E^2G\Sigma\bm{\xi}\label{swingrandom-2}
\end{eqnarray}
where
$P_{e_i}$ is given in (\ref{swing-3}),
$E=\mbox{diag}([E_{1},...,E_{n}])$, $G=\mbox{diag}([G_{11},...,G_{nn}])$, $\Sigma=\mbox{diag}([\sigma_1,...,\sigma_n])$, and $\bm{\xi}=[\dot{W_1},...,\dot{W_n}]^T$ is a vector of independent standard Gaussian random variable. {Note that (\ref{swingrandom-1})-(\ref{swingrandom-2}) is a set of stochastic differential equations.}

Since we are interested in ambient oscillation of the system around stable steady state due to load variation, (\ref{swingrandom-1})-(\ref{swingrandom-2}) can be linearized around the steady state as the following:
\begin{eqnarray}
\left[\begin{array}{c}\dot{\bm{\delta}}\\\dot{\bm{\omega}}\end{array}\right]&=&
\left[\begin{array}{cc}{{0}}&{I_n}\\-M^{-1}\frac{\partial{\bm{P_e}}}{\partial{\bm{\delta}}}&-M^{-1}D\end{array}\right]
\left[\begin{array}{c}{\bm{\delta}}\\{\bm{\omega}}\end{array}\right]\nonumber\\
&&+\left[\begin{array}{c}0\\-M^{-1}E^2G\Sigma\end{array}\right]\bm{\xi}\label{swing-matrix}
\end{eqnarray}
Let $\bm{x}=[\bm{\delta},\bm{\omega}]^T$, $A=\left[\begin{array}{cc}{{0}}&{I_n}\\-M^{-1}\frac{\partial{\bm{P_e}}}{\partial{\bm{\delta}}}&-M^{-1}D\end{array}\right]$, $B=[0,-M^{-1}E^2G\Sigma]^T$, then (\ref{swing-matrix}) {can be represented by the following set of stochastic differential equations}:
\begin{equation}
\dot{\bm{x}}=A\bm{x}+B\bm{\xi}
\end{equation}
Specifically, $\bm{x}$ is a vector Ornstein-Uhlenbeck process \cite{Gardiner:2009}.

If the state matrix $A$ is stable, the stationary covariance matrix $C_{\bm{x}\bm{x}}=\left[\begin{array}{cc}C_{\bm{\delta}{\bm{\delta}}}&C_{\bm{\delta}{\bm{\omega}}}\\C_{\bm{\omega}{\bm{\delta}}}&C_{\bm{\omega}{\bm{\omega}}}\end{array}\right]$ can be shown to satisfy the following Lyapunov equation \cite{Hines:2015}\cite{Gardiner:2009}:
\begin{equation}
AC_{\bm{x}\bm{x}}+C_{\bm{x}\bm{x}}A^T=-BB^T \label{lyapunov}
\end{equation}
which combines nicely the statistical properties of states and the model knowledge.
This relation between the covariance matrix $C_{\bm{x}\bm{x}}$ and the system state matrix is important and has been utilized in \cite{Hines:2015} to compute the covariance matrix $C_{\bm{x}\bm{x}}$ based on the model knowledge $A$ and $B$. The main novel insight provided by this paper is that it is possible to utilize (\ref{lyapunov}) the other way round, i.e., to estimate the state matrix $A$  using measurements $C_{\bm{x}\bm{x}}$. This has profound implications for power system model estimation and stability analysis as demonstrated later in this paper.

Following the derivations provided in the appendix, we arrive at the following expressions for the Jacobian and damping matrix:
\begin{align}
\frac{\partial{\bm{P_e}}}{\partial{\bm{\delta}}}&=MC_{\bm{\omega}{\bm{\omega}}}C_{\bm{\delta}{\bm{\delta}}}^{-1}\label{rdd}\\
D_k &= \frac{1}{2}M_k^{-1} G_k^2 E_k^4 \Sigma_k^2 /[C_{\bm{\omega\omega}}]_{kk}\label{rww}
\end{align}
where $[C_{\bm{\omega\omega}}]_{kk}$ is the covariance of $\omega_k$. These relations link the measurements of stochastic variation to the generator physical model, and provide an ingenious way to estimate the Jacobian matrix and the dynamic state matrix from the measurements.

First consider (\ref{rdd}) which links the covariance matrices $C_{\bm{\delta}\bm{\delta}}$ and $C_{\bm{\omega}{\bm{\omega}}}$ with the dynamic state Jacobian matrix and the inertias $M$. Given that the inertias $M$ are typically known, and that $C_{\bm{\delta}\bm{\delta}}$, $C_{\bm{\omega}{\bm{\omega}}}$ can be estimated from PMU measurements (see Section \ref{subsectioncddcww} below), the Jacobian matrix $\frac{\partial{\bm{P_e}}}{\partial{\bm{\delta}}}$ can be easily estimated from (\ref{rdd}). Note that the network model (topology and parameters) and generators' impedances are not required to calculate the Jacobian matrix. However the methodology is not purely measurement-based as the knowledge of inertias $M$ is required.
Note that the knowledge of $\Sigma$, i.e., the standard deviations of load variations, is not necessary since it doesn't enter (\ref{rdd}). It is an advantage as it indicates that the noise intensity does not affect the performance of the method.

Calculation of the full state matrix $A$ is straightforward if the damping $D$ is known---see (\ref{A1}). Note that we are using the classical dynamic model, so $D$ must include all the damping effects included in the higher-order models. If, for whatever reasons, $D$ is not known or uncertain, we can additionally utilize (\ref{rww}) to calculate $D$. This is more complicated as it requires that, apart from knowing inertias $M$ and the covariance matrices $C_{\bm{\delta}\bm{\delta}}$ and $C_{\bm{\omega}{\bm{\omega}}}$, we need to know additionally the
emf's $E$, the conductance matrix $G$ (and therefore network topology and parameters, including generators's impedances) and the load variance matrix $\Sigma^2$. The load variance matrix can be naturally estimated from the voltage and current measurements on load substations. Load variability is directly linked to the overall power consumption, and can be also naturally inferred from a combination of historical data and state estimation \cite{Crow:2013}-\cite{Nwankpa:2000}.
More details will be discussed in Section \ref{subsectioncddcww}. Note that here we have to assume to know the network model (which was not required to estimate the Jacobian matrix), so application of (14) can be thought of as validation of the assumed damping values $D$ when the network model is known.

To summarize the proposed hybrid methodology, estimation of the Jacobian matrix requires only the knowledge of generator inertias $M$ and the covariance matrices $C_{\bm{\delta}\bm{\delta}}$ and $C_{\bm{\omega}{\bm{\omega}}}$ which can be calculated from PMU measurements. Estimation of the whole state matrix $A$ requires additionally that generator damping coefficients in $D$ are known. If $D$ is unknown or uncertain, we can estimate it if we know the network model, generator emf's and variances of load variations.

{Note that the prerequisite of the proposed methodology is that the system state matrix $A$ is stable such that (\ref{lyapunov}) holds. If unfortunately a stable limit cycle exists and hides behind the noise, we may firstly apply the technique proposed in \cite{Wangxz:2015} to diagnose if a limit cycle exists. If there is no limit cycle, then we can confidently apply the proposed method to estimate the Jacobian matrix and the system state matrix, or the damping coefficients $D$.}

\subsection{Determination of Covariance Matrices $C_{\bm{\delta}\bm{\delta}}$ and $C_{\bm{\omega}{\bm{\omega}}}$} \label{subsectioncddcww}
{ The theoretical stationary covariance
matrix is defined as:}
\begin{equation}\label{cdd}
  C_{\bm{\delta}\bm{\delta}}=\left[\begin{array}{cccc} C_{\delta_1\delta_1}&C_{\delta_1\delta_2}&\dots&C_{\delta_1\delta_n}\\
  C_{\delta_2\delta_1}&C_{\delta_2\delta_2}&\dots&C_{\delta_2\delta_n}\\
  \vdots&\vdots&\ddots&\vdots\\
  C_{\delta_n\delta_1}&C_{\delta_n\delta_2}&\dots&C_{\delta_n\delta_n}
  \end{array}\right]
\end{equation}
where $C_{\delta_i\delta_j}=\E[(\delta_i-\mu_i)(\delta_j-\mu_j)]$, and $\mu_i$ is the mean of $\delta_i$.
{Nevertheless, $C_{\bm{\delta}\bm{\delta}}$ is typically unknown in practice due to limited data. Therefore, we use the sample covariance matrix $Q_{\bm{\delta}\bm{\delta}}$ to estimate $C_{\bm{\delta}\bm{\delta}}$, each entry of which is calculated as below\cite{Gardiner:2009}:}
\begin{equation}
Q_{\delta_i \delta_j}=\frac{1}{N-1}\sum_{k=1}^N(\delta_{ki}-\bar{\delta}_i){(\delta_{kj}-\bar{\delta}_j)}\label{qdd}
\end{equation}
where $\bar{\delta}_i$ denotes the sample mean of $\delta_i$, and $N$ is the sample size. Likewise, $C_{\bm{\omega}\bm{\omega}}$ can be estimated by $Q_{\bm{\omega}\bm{\omega}}$ in the same way:
\begin{equation}
Q_{\omega_i \omega_j}=\frac{1}{N-1}\sum_{k=1}^N(\omega_{ki}-\bar{\omega}_i){(\omega_{kj}-\bar{\omega}_j)}\label{qww}
\end{equation}
A window size around $500$s is used in the examples of this paper, which shows a good accuracy.

Now, assuming that we know generator inertias $M$,
we can calculate the Jacobian matrix $\frac{\partial{\bm{P_e}}}{\partial{\bm{\delta}}}$ from (\ref{rdd}):
\begin{equation}
\left(\frac{\partial\bm{P_e}}{\partial\bm{{\delta}}}\right)=MQ_{\bm{{\omega}}{\bm{{\omega}}}}Q^{-1}_{\bm{{\delta}}{\bm{{\delta}}}}\label{approxjacobian}
\end{equation}

\subsection{The Proposed Algorithms}\label{subsectionalgorithm}
We assume that PMUs are installed at all the substations that generators are connected to, and that we can use the PMUs to calculate the values of rotor angle $\bm{\delta}$ and rotor speed $\bm{\omega}$ in steady state with ambient oscillations. Discussion how exactly it is done is beyond the scope of this paper and we refer the reader to \cite{Zhou:2011}-\cite{Liu:2011}. Further advances in measurement techniques are being made and there is a progress in adding GPS-synchronized measurements of internal machine quantities such as the field voltage and current, power system stabilizer (PSS) output, terminal voltage and current, etc. This will allow a more accurate estimation of rotor angle and frequency from the PMU measurements. Assumption about PMUs installed at all generator buses is perhaps a bit optimistic  now but quite reasonable in not too far distant future due to fast rate of deployment of PMUs in many networks around the world. Furthermore, we assume the classical generator model which is widely accepted as a reasonable approximation of generator dynamics. In simulations we will test the methodology using a higher-order model.

Assuming that the inertia parameters $M$ of generators are known,
the following algorithm provides estimation of the Jacobian matrix and the system state matrix via PMU measurements:

\begin{description}[\IEEEusemathlabelsep\IEEEsetlabelwidth{Step 1.}]
\item [\textbf{{Step 1.}}] Estimate $\bm{\delta}$ and $\bm{\omega}$ from PMU measurements.
\item [\textbf{{Step 2.}}] Calculate the sample covariance matrix $Q_{\bm{\delta}\bm{\delta}}$ and $Q_{\bm{\omega}\bm{\omega}}$ by (\ref{qdd})-(\ref{qww}).
\item [\textbf{{Step 3.}}] Estimate the Jacobian matrix $\frac{\partial{\bm{P_e}}}{\partial{\bm{\delta}}}$ by (\ref{approxjacobian}).
\item [\textbf{{Step 4.}}] If the damping $D$ of generators are known, construct the system state matrix:
\begin{equation}
  A=\left[\begin{array}{cc}{{0}}&{I_n}\\-M^{-1}\frac{\partial{\bm{P_e}}}{\partial{\bm{\delta}}}&-M^{-1}D\end{array}\right]\label{A1}
\end{equation}
\end{description}

The algorithm of estimating damping $D$ of generators requires additionally that the network conductance matrix $G$ and the variances of load variations $\Sigma^2$ are known, while $E$ can be calculated from PMU measurements. Hence we have to assume effectively that the whole system model, except the knowledge of $D$ is known. Then $D$ can be estimated from (\ref{rww}).

\section{case studies}\label{casestudy}
In this section, we present three examples to demonstrate the accuracy of the proposed algorithm.
In addition, we also investigate the influence of measurement noise and window length on the proposed algorithm. All parameters of the test systems are available online: {https://github.com/xiaozhew/Jacobian-Estimation}.

\subsection{Numerical Example I: Estimation of Jacobian matrix and Detection of Topology Change}
We consider the standard WSCC 3-generator, 9-bus system model (see e.g., \cite{Chiang:book}).
The system model considering the stochastic load variation in the center-of-inertia (COI) formulation is presented as below \cite{Crow:2013}:
\begin{eqnarray}
\dot{\tilde{\delta}}_1&=&\tilde{\omega}_1\label{9bus-1}\\
\dot{\tilde{\delta}}_2&=&\tilde{\omega}_2\\
M_1\dot{\tilde{\omega}}_1&=&P_{m_1}-P_{e_1}-\frac{M_1}{M_T}P_{coi}-D_1\tilde{\omega}_1\nonumber\\
&&-E_1^2G_{11}\sigma_1\xi_1+\frac{M_1}{M_T}\sum_{i=1}^3{E_i^2 G_{ii} \sigma_i \xi_i}\\
M_2\dot{\tilde{\omega}}_2&=&P_{m_2}-P_{e_2}-\frac{M_2}{M_T}P_{coi}-D_2\tilde{\omega}_2\nonumber\\
&&-E_2^2G_{22}\sigma_2\xi_2+\frac{M_2}{M_T}\sum_{i=1}^3{E_i^2 G_{ii} \sigma_i \xi_i} \label{9bus-2}
\end{eqnarray}
where $\delta_0=\frac{1}{M_T}\sum_{i=1}^{3}M_i\delta_i$, $\omega_0=\frac{1}{M_T}\sum_{i=1}^{3}M_i\omega_i$, $M_T=\sum_{i=1}^{3}M_i$, $\tilde{\delta}_i=\delta_i-\delta_0$, $\tilde{\omega}_i=\omega_i-\omega_0$, for $i=1,2,3$,
and
\begin{eqnarray}
P_{e_i}&=&\sum_{j=1}^{3}E_iE_j(G_{ij}\cos(\tilde{\delta}_i-\tilde{\delta}_j)+B_{ij}\sin(\tilde{\delta}_i-\tilde{\delta}_j))\nonumber\\
P_{coi}&=&\sum_{i=1}^{3}(P_{m_i}-P_{e_i})
\end{eqnarray}
The parameter values in this examples are: $M_1=0.63$, $M_2=0.34$, $M_3=0.16$; $D_1=0.63$, $D_2=0.34$, $D_3=0.16$; $P_{m_1}=0.72$ p.u., $P_{m_2}=1.63$ p.u., $P_{m_3}=0.85$ p.u.; $E_1=1.057$ p.u., $E_2=1.050$ p.u., $E_3=1.017$ p.u..

The model-based system state matrix is as follows:
\begin{equation}
A=\left[\begin{array}{cc|cc}0&0&1&0\\0&0&0&1\\\hline\multicolumn{2}{c}{\multirow{2}{*}{J}}\vline\hspace{-0.002in}&-\frac{D_1}{M_1}&0\\&&0&-\frac{D_2}{M_2}\end{array}\right]\label{A}
\end{equation}
where $J=-M^{-1}(\frac{\partial\bm{P_e}}{\partial\bm{\tilde{\delta}}}+M\frac{1}{M_T}\frac{\partial P_{coi}}{\partial\bm{\tilde{\delta}}})$, for $i=1,2$. Let $(\frac{\partial\bm{P_e}}{\partial\bm{\tilde{\delta}}})_{coi}=\frac{\partial\bm{P_e}}{\partial\bm{\tilde{\delta}}}+M\frac{1}{M_T}\frac{\partial P_{coi}}{\partial\bm{\tilde{\delta}}}$, then we have
\begin{equation}
\small{\left(\left(\frac{\partial\bm{P_e}}{\partial\bm{\tilde{\delta}}}\right)_{coi}\right)_{ij}}=\left\{\begin{array}{l}E_iE_j(G_{ij}\sin(\tilde{\delta}_i-\tilde{\delta}_j)-B_{ij}\cos(\tilde{\delta}_i-\tilde{\delta}_j))\\
+\frac{M_i}{M_T}\frac{\partial P_{coi}}{\partial\tilde{\delta_i}} \hspace{1.2in} \mbox{if $i\not=j$}\\
\sum^{n}_{k=1}E_iE_k(G_{ik}\sin(\tilde{\delta}_i-\tilde{\delta}_k)\\
+B_{ik}\cos(\tilde{\delta}_i-\tilde{\delta}_k))+\frac{M_i}{M_T}\frac{\partial P_{coi}}{\partial\tilde{\delta_i}} \hspace{0.1in} \mbox{if $i=j$}
\end{array}\right.\label{dpeddcoi}
\end{equation}
where $\frac{\partial P_{coi}}{\partial \tilde{\delta_i}}=2\sum_{k\not=i}E_iE_kG_{ik}\sin(\tilde{\delta}_i-\tilde{\delta}_k)$.

\begin{figure}[!ht]
\centering
\begin{subfigure}[t]{0.52\linewidth}
\includegraphics[width=1.8in ,keepaspectratio=true,angle=0]{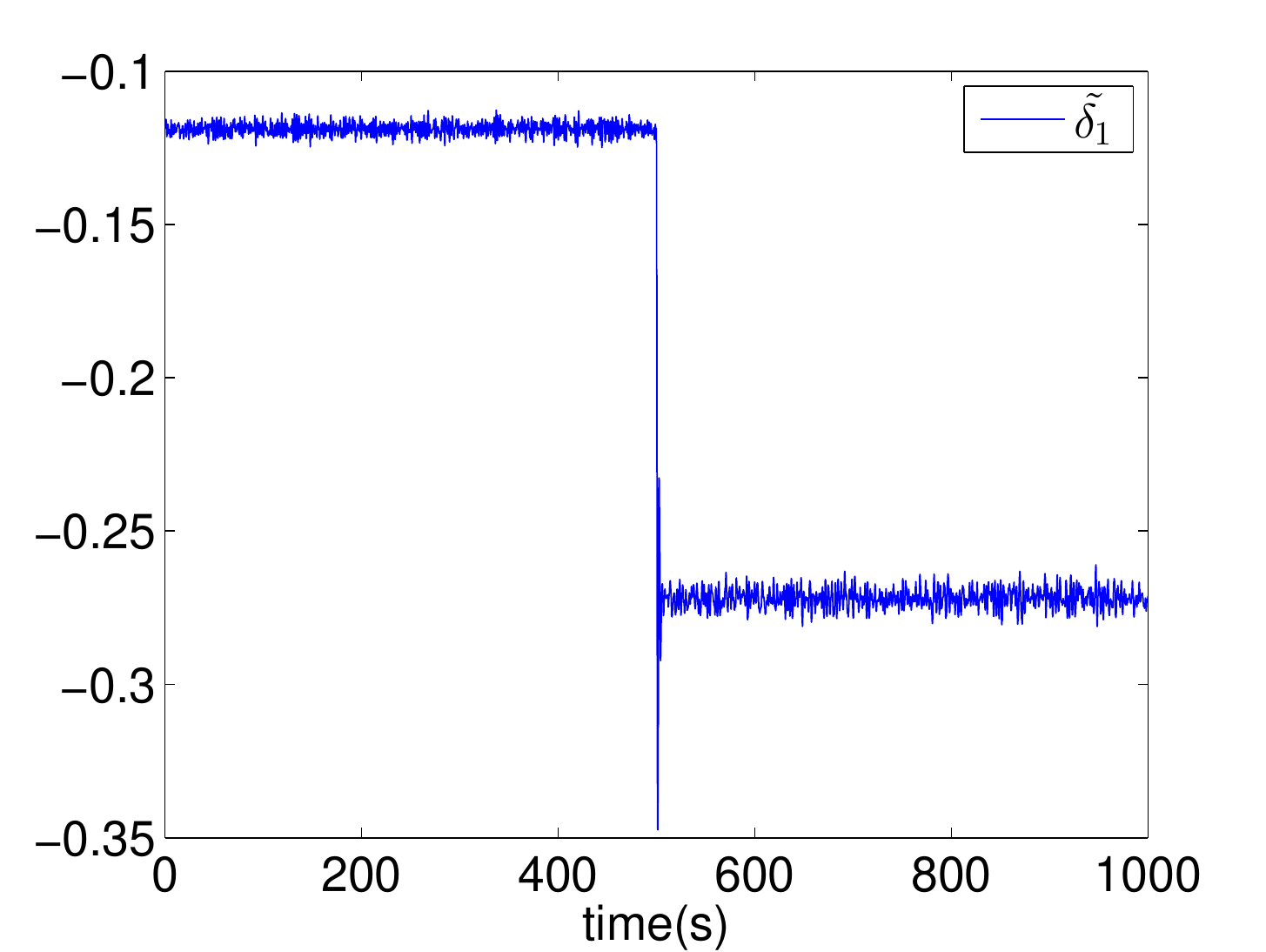}
\caption{Trajectory of $\tilde{\delta}_1$ on [0s,1000s]}\label{d1-9}
\end{subfigure}%
\begin{subfigure}[t]{0.5\linewidth}
\includegraphics[width=1.8in ,keepaspectratio=true,angle=0]{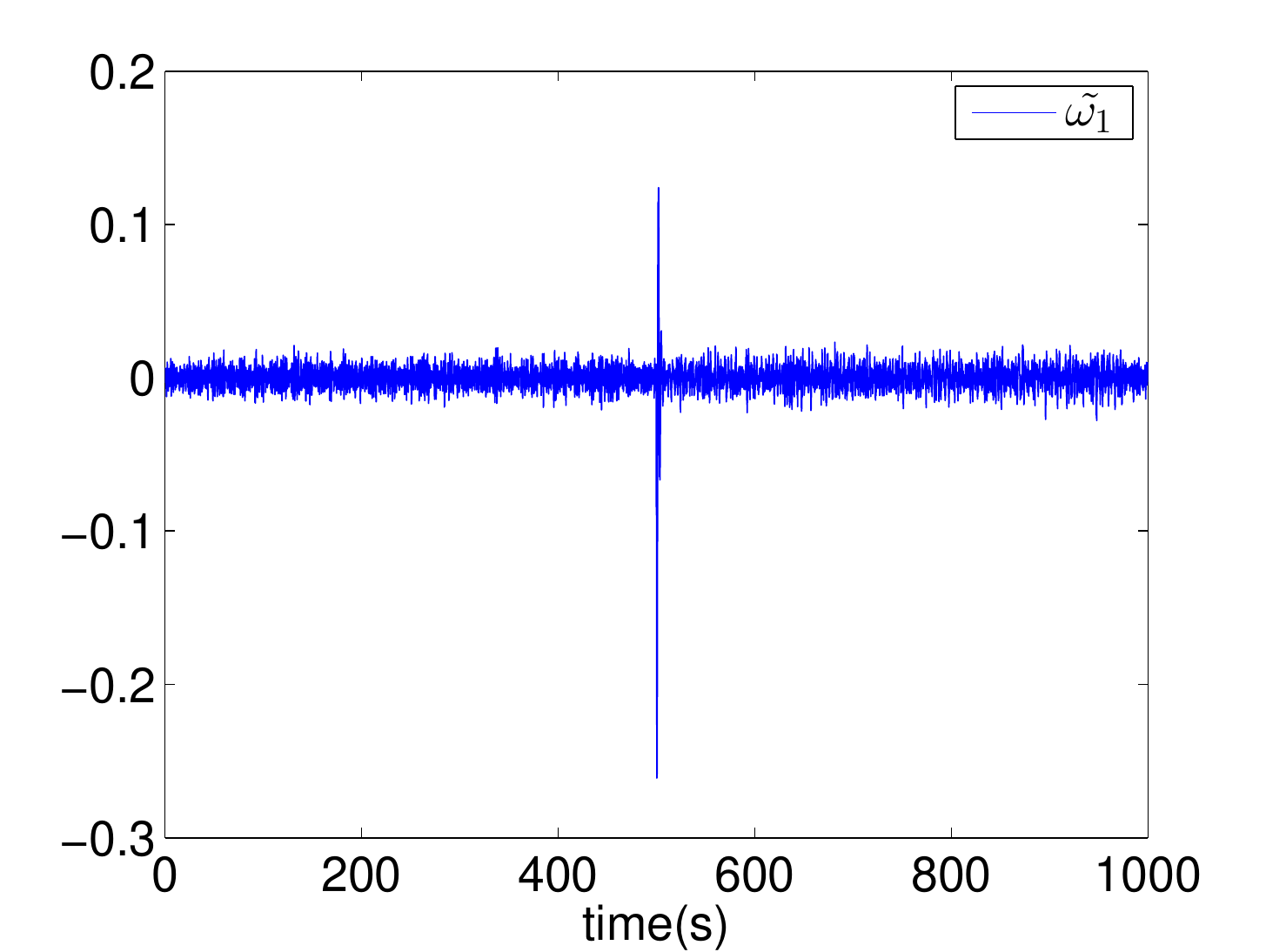}
\caption{Trajectory of $\tilde{\omega}_1$ on [0s,1000s]}\label{w1-9}
\end{subfigure}
\caption{Trajectories of $\tilde{\delta}_1$ and $\tilde{\omega}_1$ in the 9-bus system.}\label{9bus}
\end{figure}

Now we conduct the following numerical experiment to estimate the Jacobian matrix and demonstrate how the proposed methodology can be used to detect a topology change that for whatever reasons (e.g., communication errors, bad data in state estimation) went undetected and was not used to update the system model. Let $\sigma_1=\sigma_2=0.01$ p.u. denoting the standard deviations of the load variations. {The sampling rate is 10 samples per second which is much lower than the typical PMU sampling rate---48 samples per cycle, which means down-sampled PMU data is enough.} A contingency of line tripping between Bus 3 and 9 happens at 500s.  The trajectories of the rotor angle and the rotor speed of Generator 1 in COI reference are presented in Fig. \ref{9bus}, from which we see that the system is able to maintain stability after the contingency, and the state variables are always fluctuating around the nominal states due to load variations.

We first consider the situation before the contingency. If there is no load variation, i.e., $\sigma_1=\sigma_2=0$ p.u., $(\frac{\partial\bm{P_e}}{\partial\bm{\tilde{\delta}}})_{coi}$ is a model-based constant matrix that can be easily obtained from (\ref{dpeddcoi}):
\begin{equation}
\left(\frac{\partial\bm{P_e}}{\partial\bm{\tilde{\delta}}}\right)_{coi}=\left[ \begin{array}{cc} 8.053 & 1.240\\ 2.802 & 5.085\end{array}\right]\label{det_before}
\end{equation}
As there is load variation, $\bm{\delta}$ is fluctuating around its nominal state, and so is $(\frac{\partial\bm{P_e}}{\partial\bm{\tilde{\delta}}})_{coi}$. We will show that the matrix obtained from (\ref{approxjacobian}) utilizing the measurements and generator inertias only is close to  the model-based deterministic matrix (\ref{det_before}).

$Q_{\bm{\tilde{\omega}}{\bm{\tilde{\omega}}}}$ and $Q_{\bm{\tilde{\delta}}{\bm{\tilde{\delta}}}}$ before the contingency are firstly calculated from the system trajectories on $[0, 500]$s:
\begin{eqnarray}
Q_{\bm{\tilde{\delta}}{\bm{\tilde{\delta}}}}&=&10^{-5}\times\left[ \begin{array}{cc}  0.355 & -0.512\\ -0.512 &0.917\end{array}\right]\nonumber\\
Q_{\bm{\tilde{\omega}}{\bm{\tilde{\omega}}}}&=&10^{-4}\times\left[ \begin{array}{cc}  0.355 & -0.477\\ -0.477&0.967\end{array}\right]\nonumber
\end{eqnarray}
and then $(\frac{\partial\bm{P_e}}{\partial\bm{\tilde{\delta}}})_{coi}$ is computed in \textbf{Step 3} by (\ref{approxjacobian}):
\begin{equation}
\left(\frac{\partial\bm{P_e}}{\partial\bm{\tilde{\delta}}}\right)_{coi}^{\star}=\left[ \begin{array}{cc} 7.960 & 1.180\\  3.047 &  5.280 \end{array}\right]
\end{equation}
where $^\star$ denotes the Jacobian matrix estimated by the proposed method. It is observed that $(\frac{\partial\bm{P_e}}{\partial\bm{\tilde{\delta}}})_{coi}^{\star}$ and $(\frac{\partial\bm{P_e}}{\partial\bm{\tilde{\delta}}})_{coi}$ are close to each other. Specifically, the estimation error, i.e., the normalized distance between the two matrices, is:
\begin{equation}
\frac{\|(\frac{\partial\bm{P_e}}{\partial\bm{\tilde{\delta}}})_{coi}^\star-(\frac{\partial\bm{P_e}}{\partial\bm{\tilde{\delta}}})_{coi}\|_F}{\|(\frac{\partial\bm{P_e}}{\partial\bm{\tilde{\delta}}})_{coi}\|_F}=3.32\%\label{Jacobiandistance} \end{equation}
where $\|\|_F$ denotes the Frobenius norm. Assuming that $D$ is known, we can also calculate the state matrix $A$ (\textbf{Step 4}) and the resulting estimation error is:
\begin{equation}
\frac{\|A^\star-A\|_F}{\|A\|_F}=4.35\%\label{matrixdistance}
\end{equation}
Clearly, the proposed method is able to provide relatively accurate estimation for the Jacobian matrix and the system state matrix.

Next let's consider the situation after the contingency. We assume that the change of nominal states of $\bm{\delta}$ and $\bm{\omega}$ can be detected via PMU measurements, while the network topology change is undetectable. Therefore, the Jacobian matrix obtained from the model-based method via (\ref{dpeddcoi}) is:
\begin{equation}
\overline{\left(\frac{\partial\bm{P_e}}{\partial\bm{\tilde{\delta}}}\right)_{coi}^\diamond}=\left[ \begin{array}{cc} 7.338& 1.447\\ 2.831 & 4.527\end{array}\right]\label{det_after_model}
\end{equation}
where the overline denotes the value after the contingency, and $^\diamond$ denotes the value acquired from the model-based method. Indeed, this estimated Jacobian matrix deviates from the actual Jacobian matrix after the contingency shown as below:
\begin{equation}
\overline{\left(\frac{\partial\bm{P_e}}{\partial\bm{\tilde{\delta}}}\right)_{coi}}=\left[ \begin{array}{cc}  5.870 &  1.770\\ 4.001 & 4.291\end{array}\right]\label{true_after}
\end{equation}
due to the out-of-date network parameter values.

Applying the proposed algorithm, we firstly estimate the sample covariance from the trajectories on $[510,  1000]$s:
\begin{eqnarray}
\overline{Q_{\bm{\tilde{\delta}}{\bm{\tilde{\delta}}}}}&=&10^{-4}\times\left[ \begin{array}{cc}  0.0975 & -0.144\\ -0.144 &0.232\end{array}\right]\nonumber\\
\overline{Q_{\bm{\tilde{\omega}}{\bm{\tilde{\omega}}}}}&=&10^{-3}\times\left[ \begin{array}{cc}  0.0495 & -0.0676\\ -0.0676 &0.122\end{array}\right]\nonumber
\end{eqnarray}
and then compute the estimated Jacobian matrix:
\begin{equation}
\overline{\left(\frac{\partial\bm{P_e}}{\partial\bm{\tilde{\delta}}}\right)_{coi}^{\star}}=\left[ \begin{array}{cc} 6.195 & 2.031\\ 3.892 & 4.216 \end{array}\right]\label{sto_after_model}
\end{equation}
The Frobenius distance between the true (\ref{true_after}) and the estimated (\ref{sto_after_model}) Jacobian matrix is still small and is equal to $5.15\%$.
On the other hand the distance between the true (\ref{true_after}) and model-based (\ref{det_after_model}) is equal to $22.62\%$ due to assumed inaccurate network model.
As for the system state matrix $A$, the similar distances are $3.86\%$ and $21.32\%$. Those big differences between the model-based and the measurement-based matrices indicate that there was a mistake in the assumed system model.

\begin{figure}[!ht]
\centering
\includegraphics[width=3in ,keepaspectratio=true,angle=0]{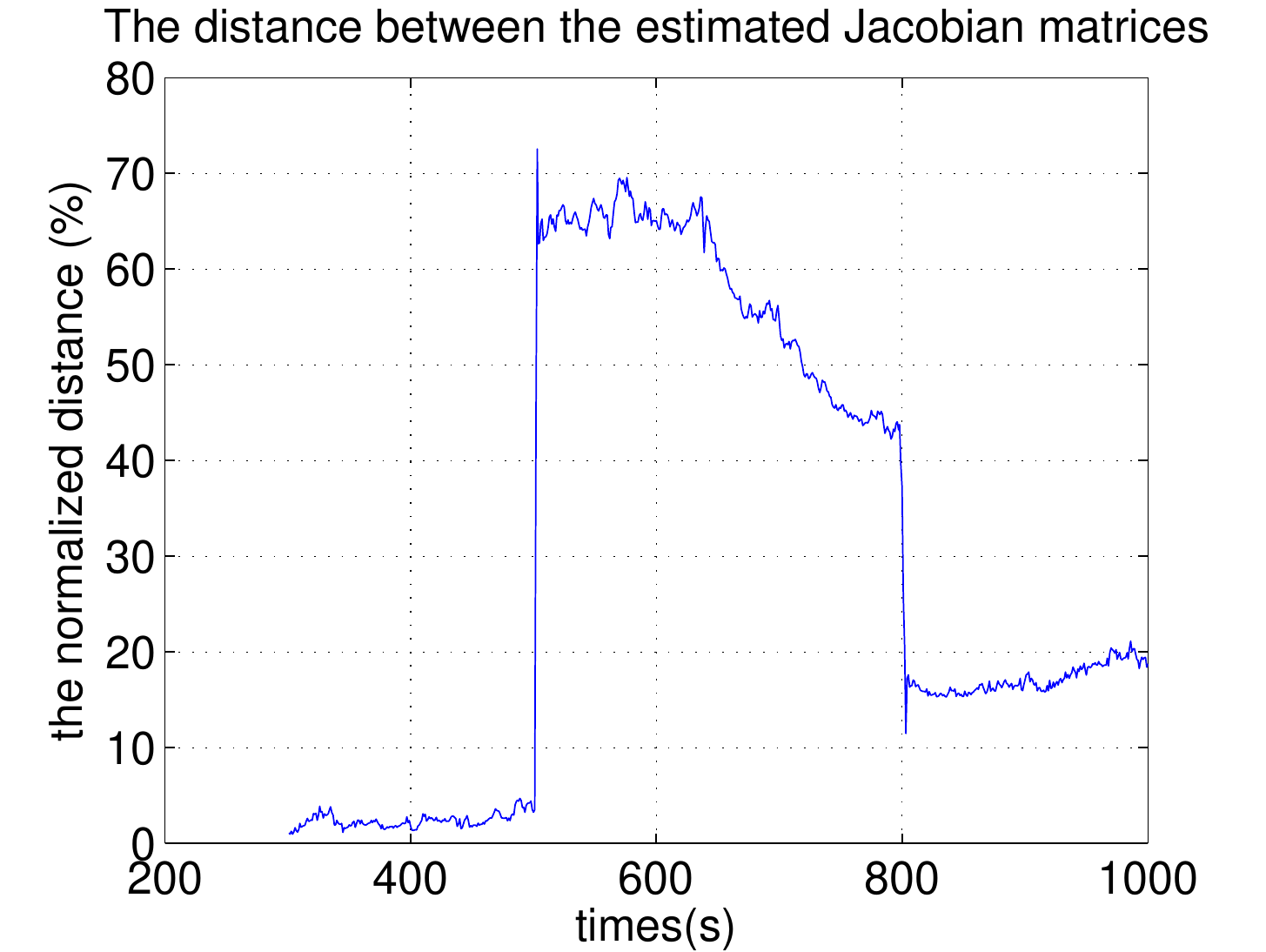}
\caption{{The normalized distance between the Jacobian matrix estimated by the proposed method using 300s moving window and the one estimated by the model-based method.}}\label{movingwindow}
\end{figure}

To make the case more realistic, a moving window of 300s is applied to the test system. The proposed algorithm is applied at every second using the past 300s data to calculate the Jacobian matrix, which is then compared with the model-based one. The normalized distance between the estimated Jacobian matrices is presented in Fig. \ref{movingwindow}. It should be noted that the proposed method only works if the system is in ambient condition, and does not work if the system is in transient condition since (\ref{lyapunov}) does not hold. Therefore, the results obtained when the moving window involving different steady-states before and after the contingency, i.e., the results between 500s and 800s, are invalid. Nevertheless, the big jump of the distance between the estimated Jacobian matrices clearly indicates that undetected variation of network topology occurs at 500s, which shows that the topology change has been detected in real time. This is further verified when the moving window fully moves to the new steady state, i.e, at 800s, and the distance between the estimated matrices still remains high.

It is also worth mentioning that both types of of network configuration errors, namely, reported network changes which have not actually occurred and unreported network changes (as shown in the above example), can be detected by the proposed method. Whenever there is a discrepancy between the assumed and the actual topology, the method is able to catch the mismatch.

Furthermore, one may wonder whether the big discrepancy between estimated matrices must imply topology change rather than other changes, e.g., load shedding. The answer is yes. If there is load shedding, it will result in large changes in $E_i$ and $\delta_i$ that will be identified by PMUs, otherwise it indicates that the assumed values of $G_{ij}$ and $B_{ij}$ and their actual values are different, which corresponds to undetected network topology change.

\color{black}

\subsection{Numerical Example II: Identification of the Location of a Topology Change}\label{subsectionexampleII}
\begin{figure}[!ht]
\centering
\includegraphics[width=3in ,keepaspectratio=true,angle=0]{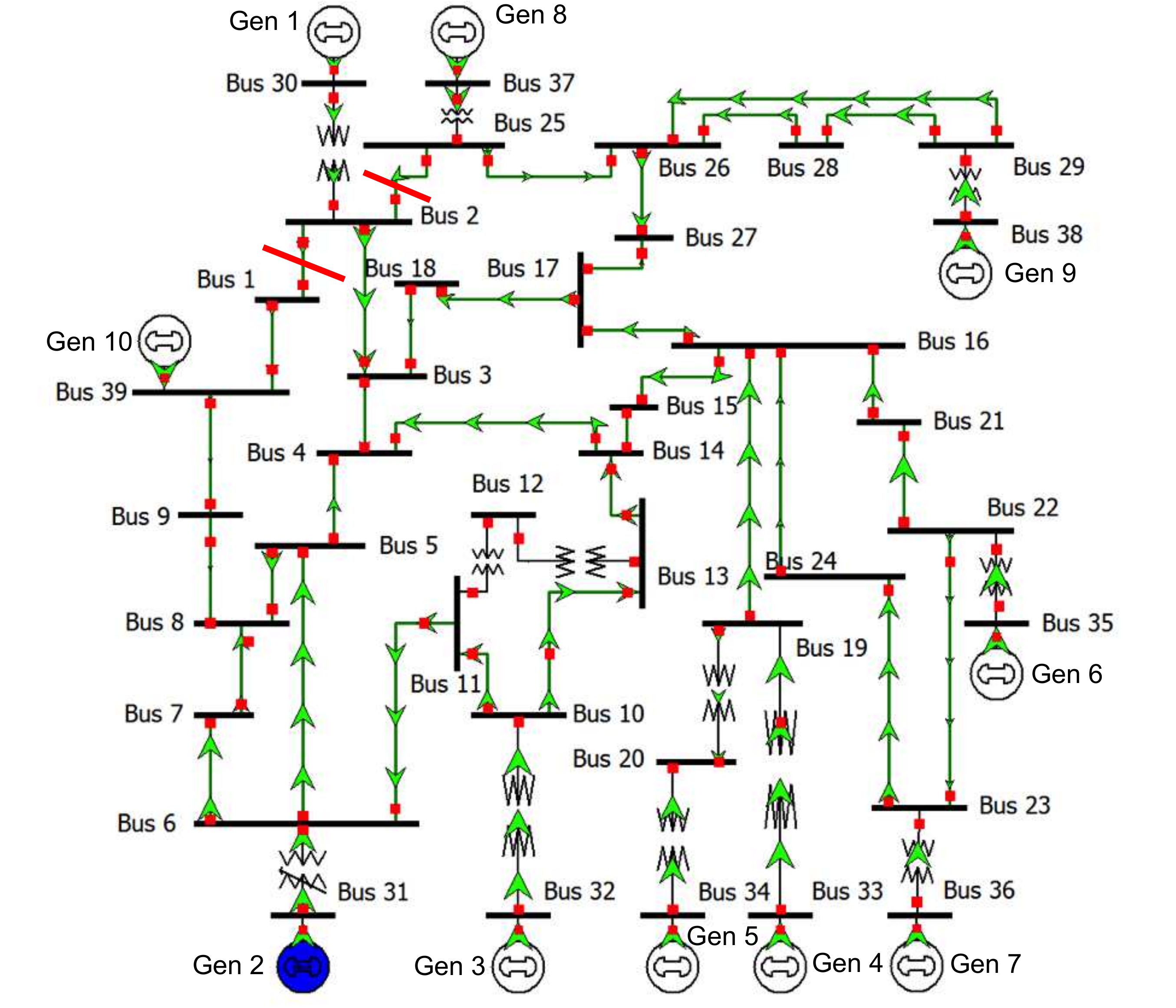}
\caption{The network topology of the IEEE 39-bus 10-generator system taken from \cite{IEEE39bus}. }\label{ieee-39}
\end{figure}

Now our aim is not only to demonstrate that the proposed methodology estimates well the Jacobian matrix when the assumed network model is inaccurate but also how to identify the source of the network model error---in this case  undetected tripping of lines.
We demonstrate this on a larger system---the IEEE 39-bus 10-generator test system, and assume that undetectable tripping of lines linking Bus 1-2 and Bus 2-25 occurs. 
For convenience, the network topology of the system is shown in Fig. \ref{ieee-39}, from which we see that the tripping lines are close to the generator at Bus 30 (Generator 1) and the one at Bus 37 (Generator 8).

Similarly as previously,
we compare the proposed method and the model-based method using normalized matrix distance in Frobenius norm. The estimation error with respect to $(\frac{\partial\bm{P_e}}{\partial\bm{\tilde{\delta}}})_{coi}$ is $18.90\%$ for the model-based method, and $4.10\%$ for the proposed hybrid method. Regarding the system state matrix $A$, if we assume that $D$ is known, the estimation error is $18.58\%$ for the model-based method, and $4.20\%$ for the proposed method. Hence the proposed method estimates well the true matrices while the model-based method gives inaccurate values due to errors in the assumed network model.

\begin{figure}[!ht]
\centering
\includegraphics[width=3in ,keepaspectratio=true,angle=0]{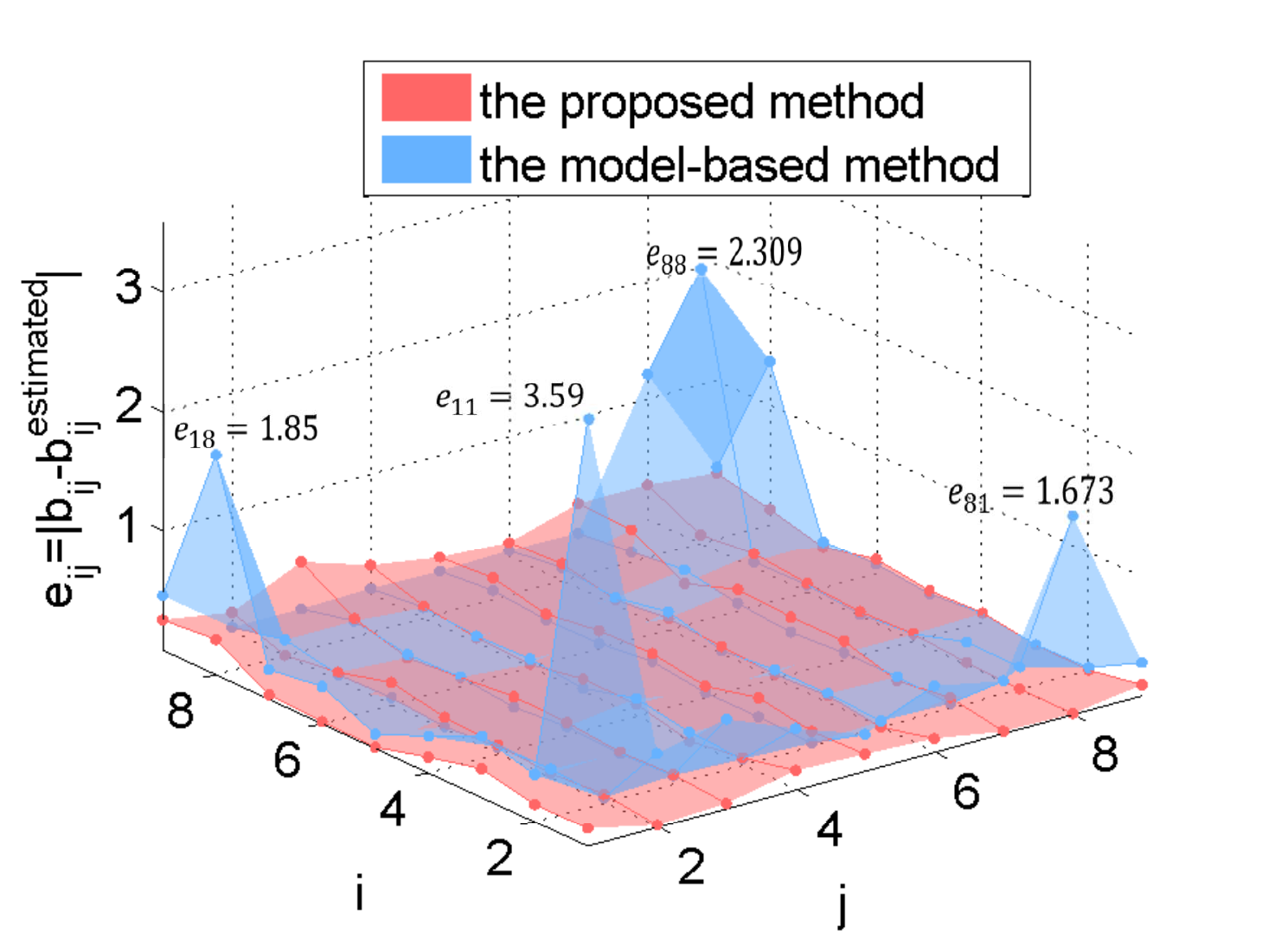}
\caption{The distance between each component of the actual Jacobian matrix and the estimated Jacobian matrices.}\label{matrixsurf}
\end{figure}

Attempting to better visualize the difference between the actual Jacobian matrix and the estimated Jacobian matrices, we use the surface plots shown in Fig. \ref{matrixsurf} to demonstrate the difference between each component of the matrices, i.e., $|b_{ij}-b_{ij}^\diamond|$ and $|b_{ij}-b_{ij}^\star|$, where $b_{ij}\in{(\frac{\partial\bm{P_e}}{\partial\bm{\tilde{\delta}}})_{coi}}$, $b_{ij}^\diamond\in{(\frac{\partial\bm{P_e}}{\partial\bm{\tilde{\delta}}})_{coi}^\diamond}$, and $b_{ij}^\star\in{(\frac{\partial\bm{P_e}}{\partial\bm{\tilde{\delta}}})_{coi}^\star}$. It is obvious from Fig. \ref{matrixsurf} that the proposed method provides more accurate approximation than the model-based method. Additionally, Fig. \ref{matrixsurf} indicates that the proposed method may not only detect the network topology change, but also help identify the location of the change. By comparing the two estimated matrices that are available in practice, we see that the biggest errors occur in $b_{11}$, $b_{18}$, $b_{81}$ and $b_{88}$ where $b_{ij}$ corresponds to $(\frac{\partial{P_{ei}}}{\partial{\tilde{\delta_j}}})_{coi}$ in the Jacobian matrix. This demonstrates that the undetected network topology change is close to Generator 1 and 8, which exactly match the fact that lines linking Bus 1-2 and Bus 2-25 trip.

The simulation results of the above two examples clearly show that the proposed method is able to provide a good approximation for the Jacobian matrix and the system state matrix by exploiting the statistical properties of the stochastic system. Additionally, the method may help detect serious discrepancies in the assumed network model and identify their sources.

\subsection{Numerical Example III: Identification of the Location of a Topology Change with Missing PMUs}

When developing the proposed methodology we have assumed that PMUs are installed at all generator buses. This is obviously a rather optimistic assumption so it is important to consider the situation when some PMUs are missing.

Assuming the PMU data at a set of Generators $K$, where $K=\{m_1,...,m_p\}$, is missing, then the proposed method can still estimate a sub-matrix of the dynamic state Jacobian matrix $\frac{\partial{\bm{P_{e}}}}{\partial{\bm{\delta}}}$, i.e., the sub-matrix $\frac{\partial{{P_{ei}}}}{\partial{{\delta_j}}}$, where $i\in\{1,...n\} \backslash K$, $j\in\{1,...n\} \backslash K$, can be estimated. More importantly, the estimated sub-matrix can still help detect the topology changes and identify the change locations.

\begin{figure}[!ht]
\centering
\includegraphics[width=3in ,keepaspectratio=true,angle=0]{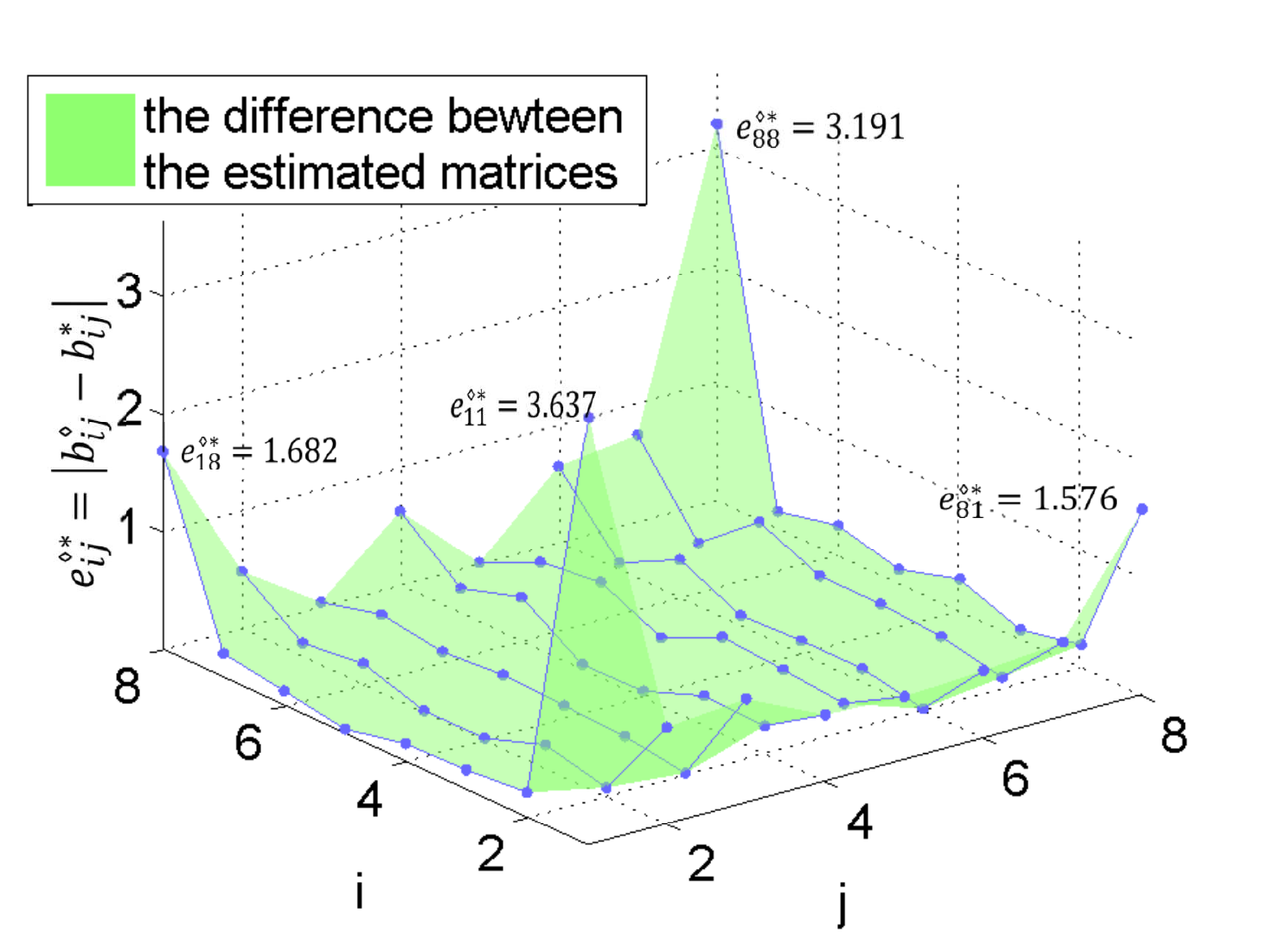}
\caption{{When the PMU at Gen 9 is missing, the distance between each component of the sub-matrix estimated by the proposed method and the one estimated by the model-based method.}}\label{matrixsurflost-diff}
\end{figure}

To show this, we revisit the IEEE 39-bus 10-generator example shown in Section \ref{subsectionexampleII}. Suppose the PMU at Generator 9 is lost, then we apply the proposed methodology to estimate the sub-matrix $\frac{\partial{{P_{ei}}}}{\partial{{\delta_j}}}$, where $i,j\in\{1,...8\}$ and $\delta_{10}$ is the reference. Similarly to the approach in Section \ref{subsectionexampleII}, we use the surface plots to show the difference between the estimated matrix by the proposed method and the one by the model-based method, i.e., $e_{ij}^{\diamond\star}=|b_{ij}^\diamond-b_{ij}^\star|$, where $^\diamond$ denotes the estimation by the model-based method and $^\star$ denotes the one by the proposed method. It's apparent from Fig. \ref{matrixsurflost-diff} that the difference between the estimated sub-matrices can still diagnose that the topology changes are close to Generator 1 and Generator 8, since the largest differences occur at $e^{\diamond\star}_{11}$, $e^{\diamond\star}_{18}$, $e^{\diamond\star}_{81}$ and $e^{\diamond\star}_{88}$ . Indeed, as long as the PMUs at Generator 1 and 8 are not missing, the proposed technique can still identify the locations of network topology changes.

\color{black}
\subsection{Impact of Measurement Noise}
The measurement noise may affect the accuracy of the proposed method. In order to investigate the impact of measurement noise, Numerical Example II is revisited in which measurement noise with standard deviation $10^{-3}$ has been added to both $\bm{\delta}$ and $\bm{\omega}$ according to the IEEE Standards \cite{IEEEStandard_amd}. Similarly, we use the normalized matrix distance to justify the accuracy of the proposed method. The estimation error of $(\frac{\partial\bm{P_e}}{\partial\bm{\tilde{\delta}}})_{coi}$ under measurement noise is $5.97\%$, and that of $A$ under measurement noise is $5.96\%$. This constitutes a small increase in the estimation error compared to the case when no noise was considered (errors of about 4\%).
It can be also shown that identification of the source of topology errors, using plots as in Fig. 3, is largely unaffected. These results show that the proposed method is still able to maintain good accuracy under measurement noise.

\subsection{Impact of Sample Size}
Like other statistical methods, the performance of the proposed method may be affected by the sample size, i.e., the window length in \textbf{Step 2} calculating the sample covariance matrices $Q_{\bm{\delta}\bm{\delta}}$ and $Q_{\bm{\omega}\bm{\omega}}$. In this part, we study the influence of the sample size on the performance of the proposed method.

\begin{figure}[!ht]
\centering
\includegraphics[width=2.8in ,keepaspectratio=true,angle=0]{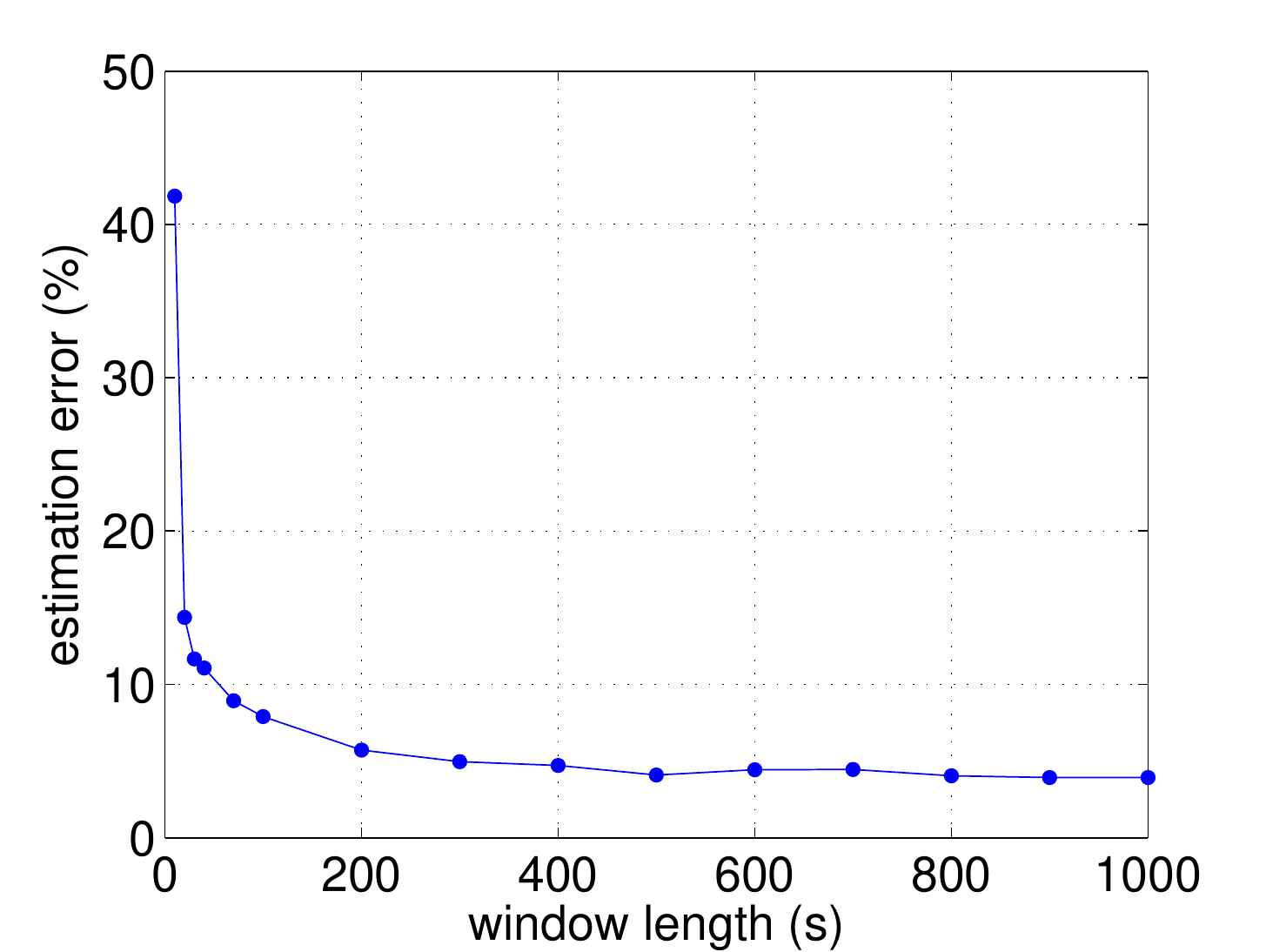}
\caption{The estimation error (\%) with respect to the window length (seconds).}\label{windowlength}
\end{figure}

For illustration purpose, we consider Numerical Example II and apply different window lengths ranging from 10s to 1000s in \textbf{Step 2}. The performance of the proposed method is justified by the normalized matrix distance between the true Jacobian matrix and the estimated Jacobian matrix shown in Fig. \ref{windowlength}. It is observed that the estimation error does not decrease substantially as the window length increases beyond 200s. The slow decrease of the error is consistent with the central limit theorem argument as the number of quasi-independent samples of covariance matrix growth linearly with the window length $t$, the error decreases as $1/\sqrt{t}$. {It should be noted that the required window length may also be affected by the quality of PMU data that we assume has been well pre-treated to highlight the intrinsic dynamics of the system.} In practice, the window length may be decided by off-line study of the system of interest.
The short window length indicates that the method is applicable in online applications. {In fact, the execution time of the proposed algorithm for 200s data is 0.023s using a computer of 2.10GHz and 8G bytes memory. The short execution time shows that the method is indeed near real-time.}

\subsection{Validation on a Higher-Order Model}

In this paper, we mainly focus on ambient oscillations around stable steady state, where the classical generator model can reasonably represent the system dynamics. Nevertheless, it can be shown that applying a higher-order generator model is equivalent to adding additional terms to the estimation which are usually small, and thus do not make a big difference.

Let us consider the generic third-order generator model:
\begin{eqnarray}
  \dot{\bm{\delta}} &=&\bm{\omega}\\
  M\dot{\bm{\omega}}&=&\bm{P_m}-\bm{P_e}-{D}{\bm{\omega}}-E^2G{\Sigma}{\xi}\\
  \dot{\bm{z}}&=&f(\bm{\delta},\bm{z})\label{higherorder}
\end{eqnarray}
where $\bm{z}$ represents the vector of additional state variables used in the third order model.
Following the same logic as in Appendix, the estimated Jacobian matrix can be represented as an expansion of power series in $C_{\bm{{\delta}}{\bm{{z}}}}$,
where $C_{\bm{{\delta}}{\bm{{z}}}}$ is typically small since: (a) $\bm{z}$ represents emf's in higher-order models and emf's do not change much in ambient conditions; and (b) angles and voltages are nearly decoupled. Hence, higher-order terms can be neglected and the estimation ${(\frac{\partial\bm{P_e}}{\partial\bm{{\delta}}})^\star}=MC_{\bm{{\omega}}{\bm{{\omega}}}}C^{-1}_{\bm{{\delta}}{\bm{{\delta}}}}$ is still accurate.

For numerical illustration,  WSCC 3-generator, 9-bus system is reconsidered.
All the generators are third-order models, each of which is also controlled by automatic voltage regulator. Simulation was done in PSAT-2.1.8 \cite{Milano:PSAT}.
Suppose $Q_{{\bm{\delta}}{\bm{\delta}}}$ and $Q_{{\bm{\omega}}{\bm{\omega}}}$ are calculated from 200s PMUs measurements, then the estimated Jacobian matrix is:
\begin{equation}
{\left(\frac{\partial\bm{P_e}}{\partial\bm{{\delta}}}\right)^{\star}}=\left[ \begin{array}{cc} 2.251 & -0.748\\-0.804 & 1.695 \end{array}\right]
\end{equation}
and the actual Jacobian matrix is:
\begin{equation}
{\left(\frac{\partial\bm{P_e}}{\partial\bm{{\delta}}}\right)}=\left[ \begin{array}{cc} 2.285 & -0.794\\-0.805 & 1.838 \end{array}\right]
\end{equation}
Thus the estimation error in terms of normalized matrix distance is $4.93\%$. Regarding the system state matrix, the estimation error is $1.04\%$. We see that the proposed method still works well for higher-order models.

\subsection{Estimation of Damping $D$}
If the damping $D$ is known, then the state matrix can be easily calculated in \textbf{Step 4} of the algorithm. In this section, we show application of the proposed method to estimation of damping $D$ under the assumption that the network model and load variation is known. Note that the knowledge of the network model was not required to estimate the Jacobian matrix, so estimation of $D$ can be thought of as a task on its own, not connected to estimation of the Jacobian matrix. The main application will be validation of the assumed model-based values of $D$ and identification and detection of any big errors in $D$.

We consider the IEEE 39-bus 10-generator test system. We assume that the standard deviation of the load variations is known and is equal to $\sigma_i=0.01$ p.u..  A comparison between the actual damping values and the estimated damping values is given in Table \ref{39-dampingtable}. Clearly the proposed method can provide good estimation for the actual damping values.

\begin{table}[!ht]
\centering
\caption{A comparison between the actual and estimated damping values}\label{39-dampingtable}
\begin{tabular}{|c|c|c|c|}
\hline
&actual damping&estimated damping&error\\
\hline
Gen 1&11.88&12.25&3.08\%\\
\hline
Gen 2&8.57&8.38&2.20\%\\
\hline
Gen 3&10.13&10.02&1.07\%\\
\hline
Gen 4&8.09&7.78&3.92\%\\
\hline
Gen 5&7.36&7.89&7.20\%\\
\hline
Gen 6&9.85&10.34&5.01\%\\
\hline
Gen 7&7.47&7.92&6.06\%\\
\hline
Gen 8&6.88&6.79&1.20\%\\
\hline
Gen 9&14.64&13.61&7.05\%\\
\hline
Gen 10&21.22&20.34&4.17\%\\
\hline
\end{tabular}
\end{table}

\section{application to online stability analysis}\label{sectionapplications}
The system state matrix $A$ provides uttermost important information on system conditions and dynamics that can be utilized in various ways. For instance, the critical eigenvalue of the state matrix can be used as a measure of proximity to instability as discussed in extensive literature (e.g., \cite{Cutsem:book}\cite{Bialek:book}). Particularly, the eigenvectors of the critical eigenvalue provide valuable information on the nature of the bifurcation, the response of the system and the control design \cite{Cutsem:book}.

\begin{figure}[!ht]
\centering
\includegraphics[width=2.8in ,keepaspectratio=true,angle=0]{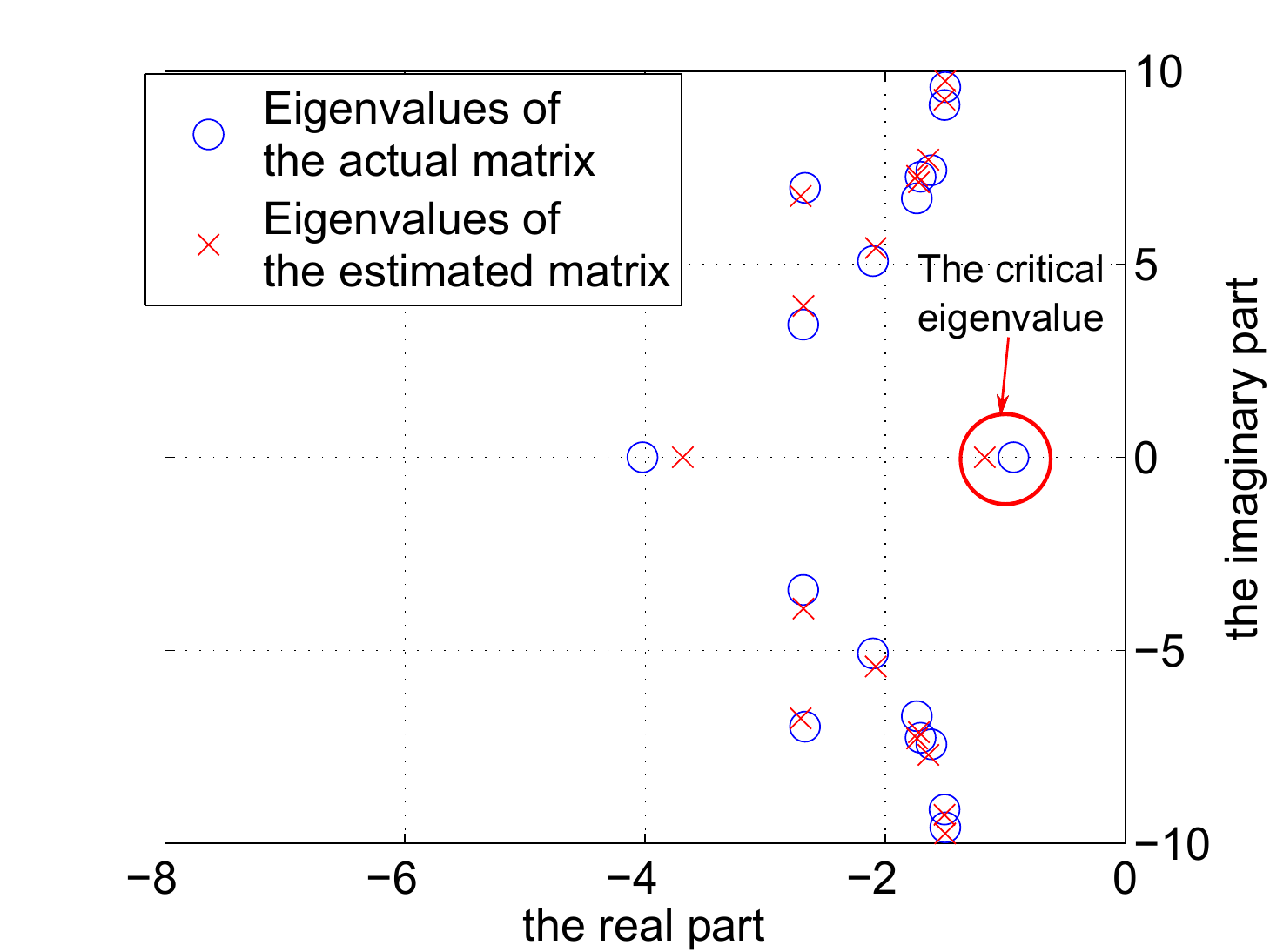}
\caption{A comparison between the eigenvalues of $A$ and those of $A^\star$.}\label{eigenpostcompare}
\end{figure}

If the damping values $D$ are known, or {have been estimated using the proposed algorithm}, we can further construct the system state matrix $A$ using (\ref{A1}) to investigate the dynamic stability of the system. We consider the IEEE 39-bus 10-generator test system. This system is similar to Numerical Example II, yet the tripping of lines occurs at Bus 2-25 and Bus 1-39 such that the system is pushed closer to the stability boundary after the contingency. A comparison between the eigenvalues of the actual system state matrix $A$ after the fault, and those of the estimated system state matrix $A^\star$ after the fault by the proposed method is shown in Fig. \ref{eigenpostcompare}, from which we see that the critical eigenvalue is well approximated and thus dynamic stability of the system can be accurately obtained.
Note that in this particular case the critical eigenvalue is real indicating an aperiodic response. This shows another advantage of the proposed methodology as {purely measurement-based} methods of mode identification (such as subspace methods, frequency domain analysis, etc.) would fail to identify non-oscillatory or high-damping modes. {In the proposed hybrid method, the ability to detect modes weakly present in measurement is enhanced by reliance on model information.}

It should be noted that in addition to a simple stability indicator, the right and left eigenvector of the critical eigenvalue may be calculated to predict the response of the system after the bifurcation and design emergency control such as generation re-dispatch. Also participation factors can be calculated. We omit those calculations here as they are straightforward \cite{Bialek:book}\cite{Dobson:1992}\cite{Dobson:2015} once the system state matrix $A$ has been determined.

\section{conclusions and perspectives}\label{sectionconclusion}

In this paper, we have proposed a hybrid measurement- and model-based method for estimating the dynamic state Jacobian matrix and the dynamic state matrix in near real-time. The proposed hybrid method works as a grey box bridging the measurement and the model, and is able to provide good estimation for the Jacobian matrix and system state matrix. As the Jacobian matrix estimation does not rely on knowledge of the network parameters, it may help detect big discrepancies and further locate the errors in the assumed network model.
In addition, an application of the estimated system state matrix in online stability monitoring is also presented. In the future, {we intend to investigate the performance of proposed method considering  higher-order models, renewable generators, lines with high series resistances, non-symmetric Y-bus cases, series/shunt compensation applications, etc.} We also plan to explore other applications of the estimated system state matrix in online oscillation analysis, model validation, power system operation such as congestion relief, economic dispatch and preventive control design.

The paper has presented mathematical foundations for the methodology verified by simulation tests. Obviously we also plan to use in future actual, rather than simulated, PMU data to validate the proposed methodology in practice. This is likely to unveil many additional challenges and hence it will be presented in a separate paper.

\appendix{}
Assuming the structure of the matrices $A$, $B$ and $C$ as defined in section \ref{sectionmodel}, the Lyapunov equations can be rewritten in the following form:
\begin{align}
    &C_{\bm{\delta}\bm{\omega}} + C_{\bm{\delta}\bm{\omega}}^T = 0 \\
    &C_{\bm{\omega}\bm{\omega}} - M^{-1}( J C_{\bm{\delta}\bm{\delta}} + D C_{\bm{\delta}\bm{\omega}}^T) = 0 \label{eq:offdiag}\\
    &M^{-1}( J C_{\bm{\delta}\bm{\omega}} + D C_{\bm{\omega}\bm{\omega}}) + \nonumber\\
    &( C_{\bm{\delta}\bm{\omega}}^T J^T {+}\color{black} C_{\bm{\omega}\bm{\omega}} D)  M^{-1} = M^{-2}G^2 E^4 \Sigma^2 \label{eq:lowdiag}
\end{align}
where, for brevity we use $J = \partial \bm{P}_e/\partial \bm{\delta}$. To express the unknown matrices $D$ and $J$ via the covariance matrix $C$ we first calculate $J$ from \eqref{eq:offdiag} arriving at
\begin{equation}
 J = M C_{\bm{\omega\omega}}C_{\bm{\delta\delta}}^{-1} + D C_{\bm{\delta\omega}}C_{\bm{\delta\delta}}^{-1} \label{eq:Jequation}
\end{equation}
Plugging this expression into \eqref{eq:lowdiag} one obtains
\begin{align}
 &M^{-1}D \hat{C}_{\bm{\omega\omega}}+ C_{\bm{\omega\omega}}C^{-1}_{\bm{\delta\delta}} C_{\bm{\delta\omega}}  +\nonumber \\
 &\hat{C}_{\bm{\omega\omega}} M^{-1}D - C_{\bm{\delta\omega}}C^{-1}_{\bm{\delta\delta}} C_{\bm{\omega\omega}} = M^{-2}G^2 E^4 \Sigma^2 \label{eq:Dequation}
\end{align}
where $\hat{C}_{\bm{\omega\omega}} = C_{\bm{\omega\omega}} +\color{black} C_{\bm{\delta\omega}}C_{\bm{\delta\delta}}^{-1}C_{\bm{\delta\omega}}$.
Given the diagonal nature of the matrix $D$, it's element on bus $k$ can be found by taking the corresponding diagonal element of both sides of \eqref{eq:Dequation} resulting in
\begin{equation}
 D_k =\frac{1}{2}\color{black}\left( M_k^{-1} G_k^2 E_k^4 \Sigma_k^2 +\color{black} M_k [R]_{kk}\right)/[\hat{C}_{\bm{\omega\omega}}]_{kk}
\end{equation}
where $R = C_{\bm{\delta\omega}}C^{-1}_{\bm{\delta\delta}} C_{\bm{\omega\omega}}- C_{\bm{\omega\omega}}C^{-1}_{\bm{\delta\delta}} C_{\bm{\delta\omega}}$ and $[X]_{kk}$ \color{black} refers to the $kk$ element of matrix $X$\color{black}. Substitution of this expression back into \eqref{eq:Jequation} results in a closed-form expression for the matrix $J$.

Within power system dynamics context, the equations can be further simplified, as the contributions from the matrix $C_{\bm{\delta\omega}}$ are typically negligible. In this case $R = 0$ and $\hat{C}_{\bm{\omega\omega}} = C_{\bm{\omega\omega}}$, so the resulting expressions take the following simple form:
\begin{align}
 J &= M C_{\bm{\omega\omega}}C_{\bm{\delta\delta}}^{-1}\\
 D_k &=\frac{1}{2}M_k^{-1} G_k^2 E_k^4 \Sigma_k^2 /[C_{\bm{\omega\omega}}]_{kk}
\end{align}

\begin{IEEEbiography}[{\includegraphics[width=1in,height=1.25in,clip,keepaspectratio,angle=270]{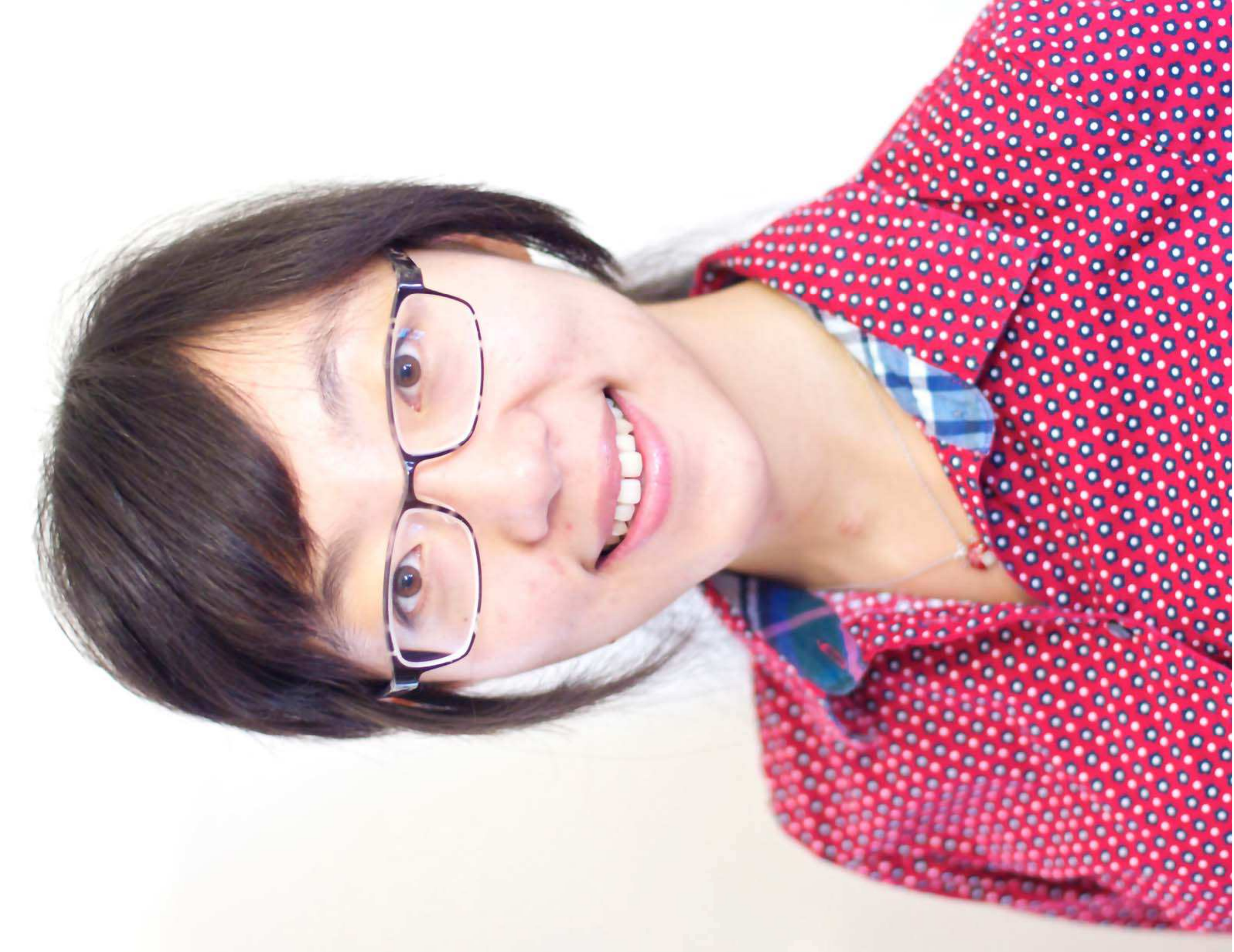}}]
{Xiaozhe Wang}
is currently an Assistant Professor in the department of Electrical and Computer Engineering at McGill University. She received the Ph.D. degree in the School of Electrical and Computer Engineering from Cornell University, Ithaca, NY, USA, in 2015, and the B.S. degree in Information Science \& Electronic Engineering from Zhejiang University, Zhejiang, China, in 2010. Her research interests are in the general areas of power system stability and control, nonlinear systems and computations.
\end{IEEEbiography}

\begin{IEEEbiography}[{\includegraphics[width=1in,height=1.25in,clip,keepaspectratio,angle=0]{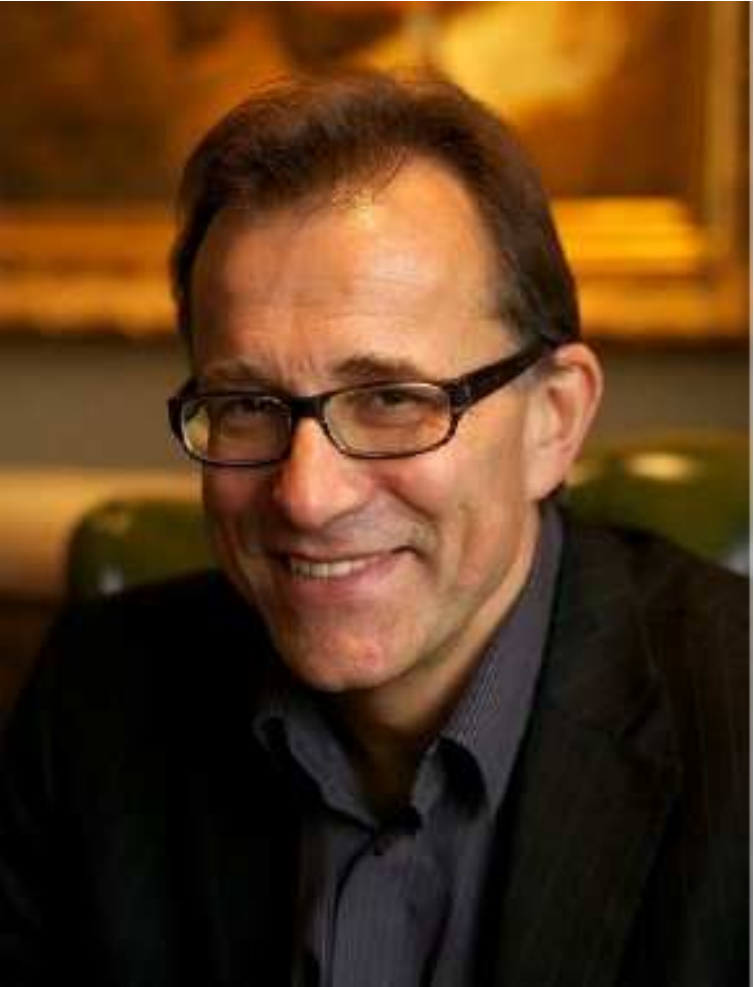}}]
{Janusz W. Bialek (F'11)} received the M.Eng. and Ph.D. degrees from Warsaw University of Technology, Warsaw, Poland, in 1977 and 1981, respectively. He is currently the Director of the Center for Energy Systems at Skolkovo Institute of Science and Technology (Skoltech) in Moscow, Russia, having previously held Chair of Electrical Engineering at The University of Edinburgh (2003-2009) and DONG Chair of Renewable Energy at Durham University (2009-2014). He has co-authored 2 books and over 100 technical papers.
\end{IEEEbiography}

\begin{IEEEbiography}[{\includegraphics[width=1in,height=1.25in,clip,keepaspectratio,angle=0]{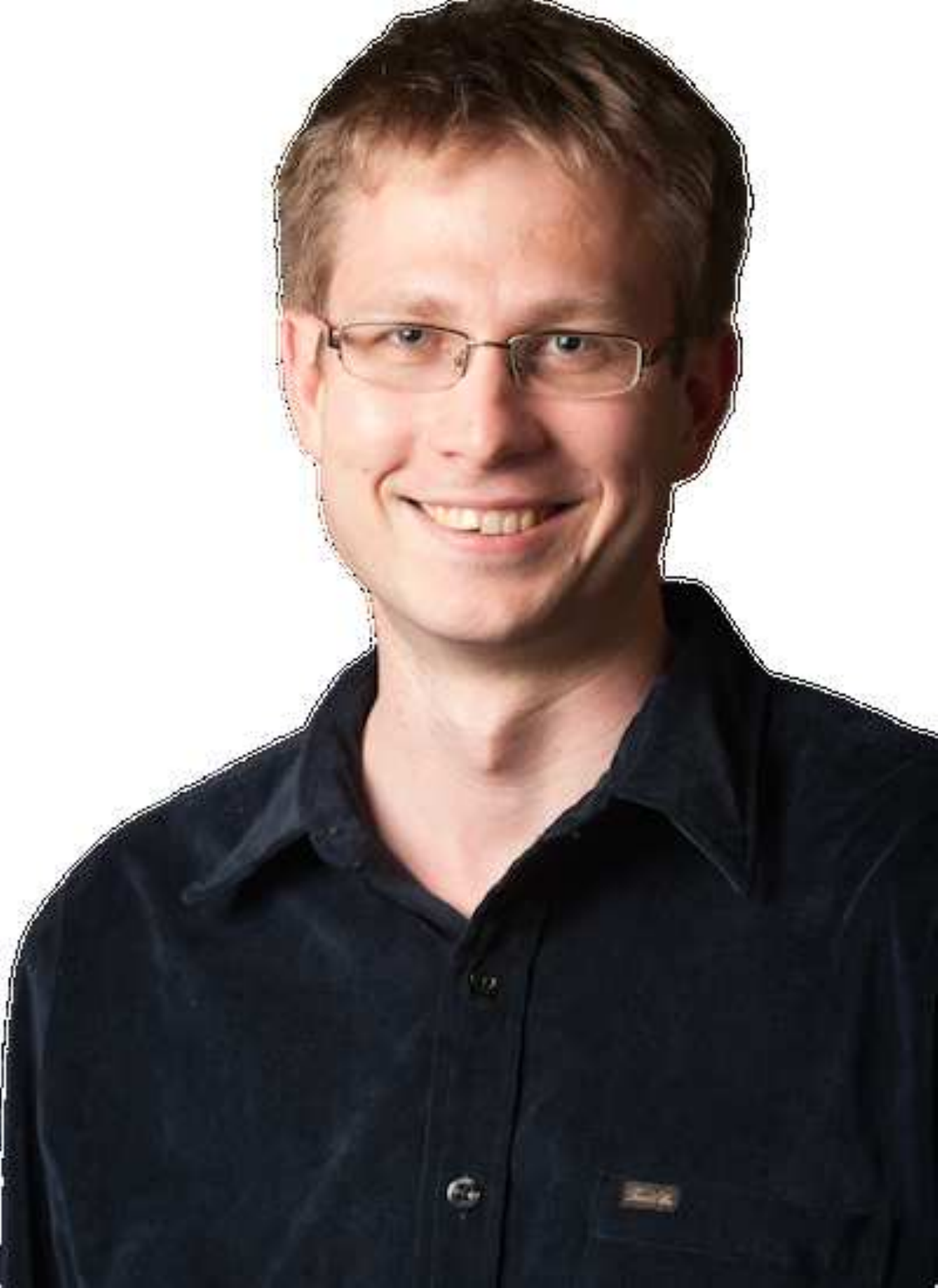}}]
{Konstantin Turitsyn} is an Associate Professor in the department of Mechanical Engineering at MIT. Before joining MIT he received PhD from Landau Institute for Theoretical Physics in 2007, and held an Oppenheimer fellow position in Los Alamos National Laboratory from 2009 and 2011. His research interests include a broad range of problems involving nonlinear and stochastic dynamics of large-scale systems.
\end{IEEEbiography}

\end{document}